\documentclass[%
 reprint,
 amSm-Ath,amssymb,
 aps,
pre,
prl]{revtex4-2}

\usepackage{graphicx}
\usepackage{dcolumn}
\usepackage{bm}
\usepackage{xcolor,soul}

\usepackage{amssymb,amsthm,amsmath}
\usepackage{xcolor,paralist,hyperref,titlesec,fancyhdr,etoolbox}

\begin{document}


\title{Cells around the corner}

\author{Aniruddh Murali\textit{$^{1}$}}
\author{Prasoon Awasthi\textit{$^{1}$}}
\author{Kirsten Endresen\textit{$^{2}$}}
\author{Arkadiusz Goszczak\textit{$^{3}$}}
\author{Francesca Serra\textit{$^{1,2}$}}%
\email{Email: serra@sdu.dk}
 
\affiliation{%
 \textit{$^{1}$}Dept. Physics, Chemistry and Pharmacy, University of Southern Denmark, Odense, Denmark. \\
 \textit{$^{2}$} Dept. Physics and Astronomy, Johns Hopkins University, Baltimore, USA.\\
 \textit{$^{3}$}~Mads Clausen Institute, University of Southern Denmark, Sønderborg, Denmark }

%




\date{\today}

\begin{abstract}
The study of spindle-like cells as nematic liquid crystals has led to remarkable insights in the understanding of tissue organization and morphogenesis. In the characterization of this anomalous liquid crystal material, we focus on the energetic cost of splay and bend deformations, in order to determine the elastic anisotropy of the material, i.e. the ratio of the elastic constants associated with splay and bend. We explore the behavior of monolayers of cells in proximity to corners, where cells arrange in splay or bend configuration, strongly dependent on the amplitude of the wedge angle. The angle at which splay and bend deformations are equally likely is determined by the ratio between splay and bend elastic constants.  We also show that the splay and bend deformations under confinement can be well approximated using equilibrium liquid crystal theory and statistical mechanics. Finally, our data suggest that for fibroblast cells the common approximation of equal bend and splay constant is valid.
\end{abstract}

\maketitle


\section{Introduction}

Living cells and tissues are examples of active matter systems, where cells generate and consume energy locally, leading to processes such as motion, shape transitions, and division. This area of research has gained significant attention, especially the emergent collective behavior and large-scale organization that extends far beyond the scale of individual units. Such collective phenomena are observed across various biological scales, from subcellular levels, such as protein aggregation within cells \cite{Zwicker2016}, to cellular processes like morphogenesis \cite{Balasubramaniam2022}, collective migration and wound healing \cite{Banerjee2019}, and even to larger-scale processes, including organ formation and the coordinated movements of bird flocks or fish schools \cite{Toner&Tu1995, Ramaswamy2010, Marchetti2013, Gompper2020}. 
The study of nematic liquid crystal-like order in cell monolayers dates back to the pioneering works of Elsdale, Gruler, and Bouligand, who observed this phenomenon in fibroblast, amoeboid, and chitin cells \cite{Elsdale1968, Gruler1999, Bouligand2008}. In recent years, the study of experimental systems exhibiting nematic-ordered regions such as bacterial cells \cite{Volfson2008, Copenhagen2020} and anisotropic cells like fibroblasts and myoblasts \cite{Duclos2014, Duclos2016} has sparked renewed interest in the role of liquid crystallinity in biology. Several experiments suggest that this nematic alignment is important in cell migration and cell-cell communication \cite{ray2017, RAY2021, kaiyrbekov2024, morales2019}, while the topological defects play a role in morphogenesis and tissue organization \cite{Kawaguchi2017, Doostmohammadi2016, Shankar2022, Saw2017, Guillamat2022, Balasubramaniam2021}. 
Experimental progress in this field has arguably been spurred by theoretical advances in the understanding of active nematic liquid crystals. Most existing theories are based on combining fluid dynamics models (Navier-Stokes equation, or a phase model) with the Q-tensor description of nematic liquid crystals \cite{giomi2011, Giomi2014, mueller2019}. In the evolution of the Q-tensor, one key ingredient is the minimization of the Frank elastic energy. Due to the already high complexity and computational cost of modeling active nematics, most studies have adopted the single elastic constant approximation \cite{Perez-Gonzales2019, Doostmohammadi2018, Blanch-Mercader2021, Turiv2020, Saw2017, Comelles2021, Duclos2016, Wang2023, Hoffmann2020} (with one notable recent exception \cite{ravnik}). This means that the constants that set the energy scale for splay and bend deformations are considered equal. However, it is well known that this is not always true. In most polymeric and lyotropic liquid crystal systems, for example, there is an order of magnitude difference between the twist elastic constant and the other two constants \cite{ferrarini1, ferrarini2, dejeu}. In bent-core nematics, the bend constant is significantly lower, and the opposite is true for nematics formed by pear-shaped molecules, which have a lower splay constant. It is reasonable to think that in active nematic systems, especially those based on bulky mesogens such as cells, the shape of the cells may influence the relative magnitude of elastic constants and the one-constant approximation may not be appropriate. A measurement of elastic anisotropy can confirm the validity of the one constant approximation. 
Toward this goal, Zhang et al. \cite{Zhang2017} have been able to determine the ratio between elastic constants in active filaments, particularly by examining F-actin in a quasi-2D environment through defect morphology analysis. The same method was successfully applied to myoblasts \cite{Blanch-Mercader2021,Mercheder2021_p2}. 

For some systems, however, there is a potential difficulty in applying this method, because the mesogen density has big spatial variation (like cyanobacteria in \cite{kurjahn2024}) or because the location of the core of the defects in cells is not clearly defined, or because cell shape changes near defects. We therefore propose an alternative method based on the analysis of cell alignment near a 2D wedge imposed by rigid walls, which allows us to control the location of the deformation.  We classify the cell alignment into bend and splay deformation, and we use corners of varying amplitude to measure the elastic anisotropy of cell layers.

\section{Results}

\subsection*{Fibroblast cells show liquid crystal-like behavior}
\begin{figure}
\centering
\includegraphics[width=\linewidth]{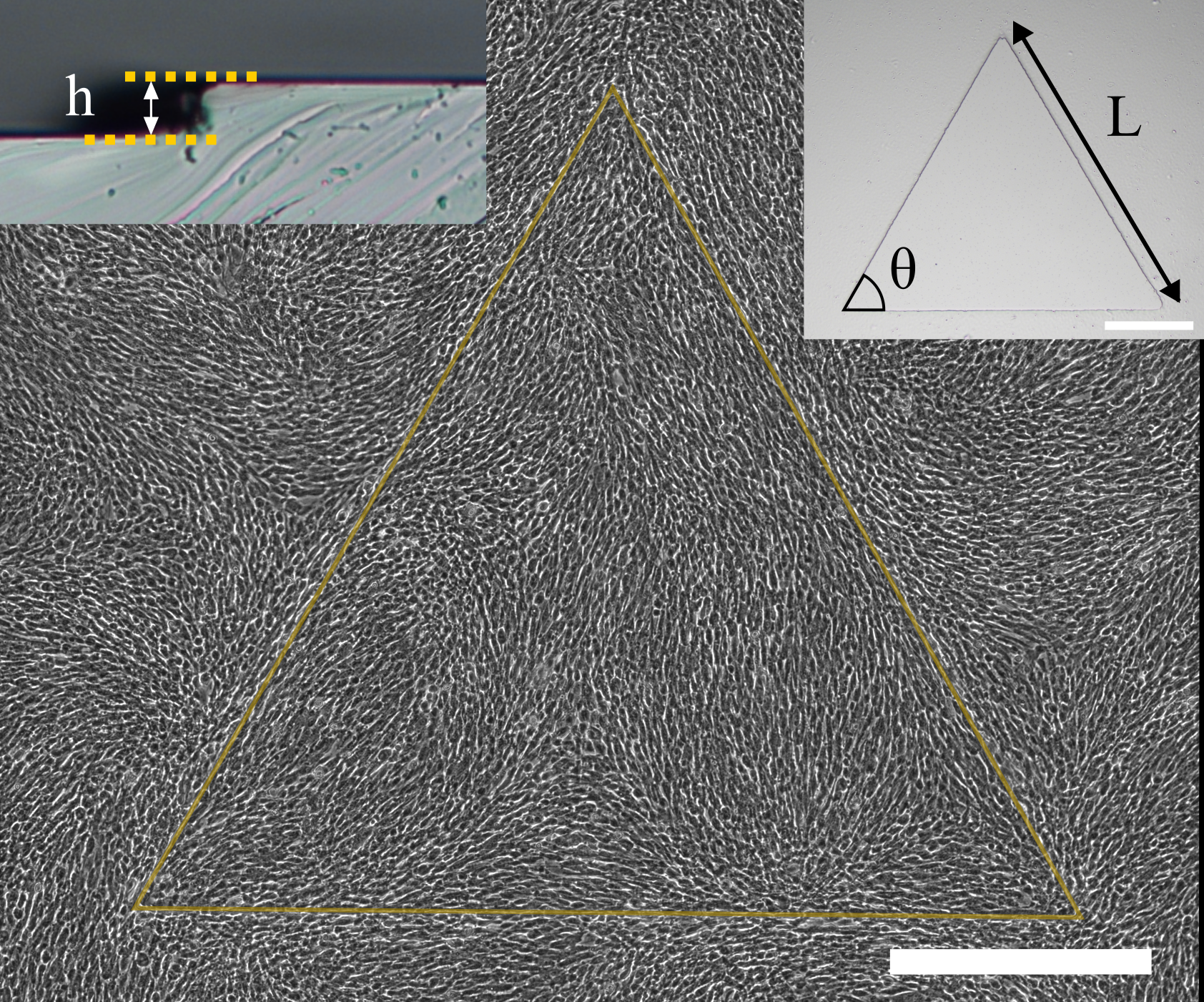}
\caption{ Phase contrast image of 3T3 fibroblast cells grown on a pillar shaped as an equilateral triangle, coated with fibronectin and imaged 48hrs after seeding. The length of the side is denoted by L (inset right), set at  $1800 \mu$m. The angle at each of the vertices denote the wedge angle given by $\theta$. The scale bar in the figure is  $500 \mu$m. The left inset shows the height of the patterned substrate relative to base PDMS (h), fixed at $10 \mu$m.
}
\label{fig:Figure 1 Main}
\end{figure}

To understand deformations in cell layers near corners, in our study, we employ substrates featuring pillars with a triangular base. As depicted in Fig.\ref{fig:Figure 1 Main} right inset, the length (L) of one side of the triangle is fixed at 1800 $\mu$m, ensuring that the distance between corners is large enough that the cell alignment near each corner is not influenced by the other corners. The height of the triangles (h) indicated in left inset in Fig.\ref{fig:Figure 1 Main}. is 10 $\mu$m to isolate the top of the pillars from the bottom of the substrate. While it is possible for cells to overcome the 10$\mu$m barrier, this height is sufficient to ensure that the cell alignment on the triangles is mostly dictated by the edges \cite{Kaiyrbekov2023} . The triangles are spaced sufficiently apart to prevent mutual influence. Additional fabrication details are provided in the Materials and Methods section. 

NIH-3T3 fibroblast cells are then cultured on these patterned substrates, with the vertex angle ($\theta$) varying from $\frac{\pi}{6}$ to $\frac{5\pi}{6}$. The fibroblast cells are seeded on fibronectin-coated substrates, and their proliferation is observed after a 48hr period (see for example Fig.\ref{fig:Figure 1 Main}, with the outline of the triangle pattern plotted in yellow). As expected, the edges of the triangles provide a strong alignment cue for cells \cite{bade2018edges}, which are even sensitive to much smaller ridges \cite{Endresen2021,Kaiyrbekov2023}. We verify that the edges of the triangles guide the alignment of the fibroblast long axis, as evident from Fig.\ref{fig:Figure 1 Main}. In addition, we verify that the deformations near the vertices of the triangles are splay-like or bend-like and can be systematically characterized into one or the other.

\begin{figure*}[th!]
\centering
\includegraphics[width=\textwidth]{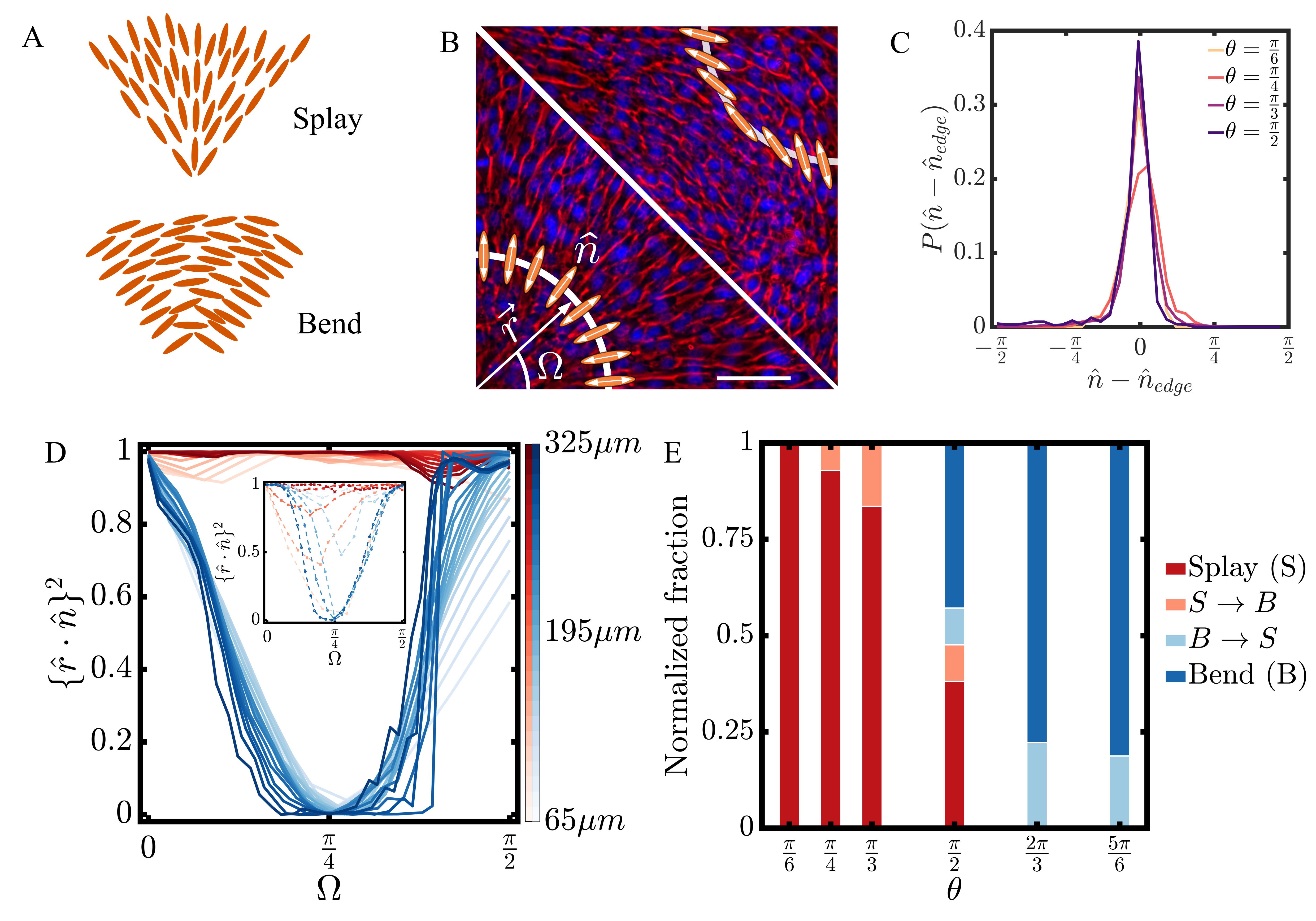}
\caption{(A) Schematic of planar splay and planar bend in liquid crystals. Each individual unit (orange) represents a liquid crystal mesogen.  (B) Example of deformation of cells near the right corner of two different isosceles right triangles. The images are placed diagonally opposite to each other and cropped to emphasize the difference. On the left bottom corner a splay deformation is shown, while the right top corner shows a bend-like deformation. The red channel is the phase contrast imaging and the blue channel is the fluorescence imaging of cell nuclei stained with NucBlue LiveCell Stain for better visualization. The images of the cells were taken 48hrs after seeding cells on the triangles. Orange ellipse are used as a proxy for a cell, with the orientation given by $\hat{n}$. $\Omega$ denotes the angular position of each cell sub-unit and $\hat{r}$ being the unit radial vector from the vertex of the triangle to the cell. (Scale Bar: 100 $\mu$m) (C) Distribution of angles between the orientation of the cell immediately next to edge and the edge director for different wedge angles $\frac{\pi}{6} \leq \theta \leq \frac{\pi}{2}$, normalized by the number of observed cells. (D) $\{\hat{r}\cdot\hat{n}\}^2$\ vs\ $\Omega$ for two experimental realizations of splay (in red) and bend (in blue) respectively with the vertex angle of $\theta=\frac{\pi}{2}$. The color intensity goes from light to dark both in splay (red) and bend (blue) as the distance from the vertex is increased from $65\mu$m to $325\mu$m. Inset $\{\hat{r}\cdot\hat{n}\}^2$ vs $\Omega$ for two experimental cases with splay to bend (blue) and bend to splay (red) is observed. (E) Fraction of observed splay and bend deformations as a function of the wedge angle $\theta$ (details in SI Table S1). Red columns are for pure splay deformation, light red for splay-to-bend transition, light blue for bend-to-splay and blue for pure bend deformation.} 
\label{fig:Figure 2 Main}
\end{figure*}

\subsection*{Splay and bend deformation}
When we classify splay-like and bend-like deformations in our systems we should keep in mind that the alignment of the cells near the edges is always parallel to the edges. While this planar alignment is compatible with a pure-splay deformation, it cannot accommodate a pure-bend deformation like the one depicted at the bottom of Fig.\ref{fig:Figure 2 Main}A. Fig.\ref{fig:Figure 2 Main}B shows an example of this: two $90^\circ$ vertices from two different isosceles right triangles are shown, 48 hrs after seeding cells at the density of 500 $\mathrm{cells/mm}^2$. The overlayed rods mirror the direction of individual cells, which can be seen in the figure in phase contrast (red) and nuclear fluorescence (blue). Both techniques show that cells adopt two characteristic deformations, i.e. splay-like (bottom left corner) or bend-like (top right corner). The splay deformation is easily recognizable, while the bend deformation differs from the ``pure bend'' in Fig.\ref{fig:Figure 2 Main}A, as it is frustrated by the triangular wedge.   

The orientation of the cell monolayer is estimated using OrientationJ \cite{Puspoki2016} at grid points ($\mathrm{13}\mu m \times \mathrm{13}\mu m$). At a point $\vec{r}$ the orientation is given by $\hat{n}$, while $\vec{r}$ subtends an angle $\Omega$ with respect to edge of the triangle. Calculating the angle between the triangle edge and the grid points close to the edge, as seen in Fig.\ref{fig:Figure 2 Main}.C, we see as expected the cells align mostly along the edges (compatibly with the typical uncertainties in measuring the orientation of cells) \cite{Duclos2016,Endresen2021,Kaiyrbekov2023,Blanch-Mercader2021}. We verify that this behavior is consistent for different values of the wedge angle $\theta$ from $\frac{\pi}{6} \, \mathrm{to} \, \frac{\pi}{2}$. 

To characterize whether the cell monolayer exhibits splay or bend deformations near the corner, we use the quantity $\{ \hat{r} \cdot \hat{n}\}^2$. For example, for a planar splay deformation near a right corner, $\{\hat{r} \cdot \hat{n}\}^2$ will equal 1 as we move from $\Omega = 0$ to $\Omega = \frac{\pi}{2}$. In contrast, for planar bend-like deformation, $\{\hat{r} \cdot \hat{n}\}^2$ equals 1 at $\Omega = 0$ and $\frac{\pi}{2}$, but drops to 0 at $\Omega = \frac{\pi}{4}$. When we plot this function for our experimental data, obtained for the right angle of two different triangles, the behavior matches the expected patterns for both planar splay and planar bend deformations, as shown in Fig.\ref{fig:Figure 2 Main}.D. This visualization can distinguish well the corners where splay deformation is dominant from those where bend deformation is dominant.
The deformations near corners influence the liquid crystal deformation further from the wedge, which is evident as we increase ${r}$ from $\mathrm{65}\,\mu$m  to $\mathrm{365}\,\mu$m, with darker shades representing greater distances in both splay (red lines) and bend (blue lines) deformations. 

There are cases where the deformation changes as one moves away from the corner and transitions from splay to bend or bend to splay. Typically, this indicates the presence of either a $+\frac{1}{2}$ or $-\frac{1}{2}$ topological defect near the corner. Examples of these transitions are shown in the inset of Fig.\ref{fig:Figure 2 Main}.D. The red dotted line illustrates a deformation near a wedge, transitioning from bend to splay as the distance increases (indicated by the darkening line), suggesting the existence of a $+\frac{1}{2}$ topological charge. Conversely, the blue line shows a transition from splay to bend as the distance from the vertex increases, indicating a $-\frac{1}{2}$ defect. Examples of these deformations in our experiments are provided in the Supplementary Information (SI) SI Fig.S1. 

Therefore, deformations at any given wedge angle can be classified as splay, bend, splay-to-bend, or bend-to-splay based on the function's profile. 

To classify a given wedge into one of the four categories, we apply specific criteria. If the function $\{ \hat{r} \cdot \hat{n}\}^2$ does not go below 0.25 for more than 3 consecutive distances then we classify the wedge to be splay-like. Similarly if the minimum of the function $\{ \hat{r} \cdot \hat{n}\}^2$ does not go over 0.75 for 3 consecutive distances then we classify it as bend-like. The transition from splay to bend occurs when the splay rule is broken and viceversa. These rules actually help us avoid errors in the categorization that can occur due to few misaligned or round cells (we further elaborate on this point in SI Fig. S2).  

Using this classification, we plot the fraction of observed deformation types at different wedge angles (Fig.\ref{fig:Figure 2 Main}.E). The data are collected from multiple experimental runs on the same day and from different samples prepared on various days to ensure that substrate preparation did not introduce any bias (see SI Table S1). As expected, at smaller wedge angles, cells predominantly arrange in a splay deformation with planar anchoring near the edges (indicated by the dark red color). Further from the vertex, as $\theta$ increases, bend deformations begin to appear, signaling the onset of splay-to-bend transitions (light red color). At $\theta = \frac{\pi}{2}$, the fractions of splay and bend (dark blue color) deformations are equal across different runs, suggesting no strong preference for one deformation over the other. Additionally, the likelihood of encountering splay-to-bend and bend-to-splay transitions (light blue color) is also similar, suggesting energy symmetries between splay and bend deformation. As the wedge angle increases further, bend deformations become more and more likely, which aligns again with intuitive expectations. We would like now to use these results to give quantitative estimates of the elastic anisotropy.

\subsection*{Deformation Energy}
In order to extract quantitative information, we write down the energy terms associated with the two deformations. Due to the low activity of our system, we do not incorporate dynamics in the description. First we need to describe the splay and bend deformations near a corner in terms of the director ${\hat{n}}$. A pure splay deformation with planar anchoring can be conceptualized as part of a $+1$ topological defect, with the vertex at the center of the defect core \cite{Chandrasekhar1986, deGennes1993}. For the bend deformation, we know that near the wedges the vector should align along the edge, while at the midpoint, the orientation must be completely perpendicular to the radial direction. We write the nematic director ${\hat{n}}$ in cylindrical coordinates and for the splay and bend deformation we get:  
\begin{equation}
\hat{n}_{splay}=\hat{r} \nonumber 
\\
\hat{n}_{bend}=cos\, \left ( \frac{\pi}{\theta} \phi\right ) \,\hat{r} - sin\, \left(\frac{\pi}{\theta} \phi\right) \,\hat{\phi} 
\label{eqn:ideal_n}
\end{equation}

The function $\{\hat{r} \cdot \hat{n}\}^2$ is plotted in SI Fig.S3 using the above expressions for ${\hat{n}}$. We can then write down the Frank Oseen energy per area $\mathbf{f_{\textit{FO}}}$ for 2-D nematic liquid crystals \cite{Kleman2003}. Since we are looking at cell monolayer where cells are confined to a 2D plane, we discard the twist energy term.

\begin{equation}
\displaystyle
 f_{FO}=\frac{1}{2}k_1^*(\nabla\cdot{\hat{n}})^2+\frac{1}{2}k_3^*({\hat{n}}\times(\nabla\times\hat{n}))^2  \label{eqn:frank_Oseen_main} 
\end{equation}

where $k_1^*$ and $k_3^*$ denote the 2D-elastic constants for splay and bend, respectively. For simplicity, the splay and bend elastic constants are referred as $k_1 \,\mathrm{and}\, k_3$ from now on, but one should keep in mind that their units differ from the usual definition. The total energy can be calculated using  

 \begin{equation}
\displaystyle
 E_{tot}=\int_A f_{FO}\,dA + 2\int \,W(\hat{n}\times\hat{\tau})\,dl  \label{eqn:totalenergy_main} 
\end{equation}

where W is the (line) anchoring energy per unit length, and $\hat{\tau}$ is the tangential to the edge. The detailed calculations are given in the SI theory section.

Having concluded from Fig.\ref{fig:Figure 2 Main}.C that the cells have strong planar alignment, we consider the contribution of anchoring energy ($W$) to be zero. To compute the total energy associated with splay deformation, we substitute the splay director ($\hat{n}_{splay}$) from \eqref{eqn:ideal_n} into  \eqref{eqn:frank_Oseen_main}. By integrating over the area, with 
$\phi$ ranging from 0 to wedge angle $\theta$ and $r$ from the defect core radius $\epsilon$ to a characteristic length $l$ away from the corner, we get 
\begin{align}   
E_{tot_{splay}} = \frac{1}{2}k_1\theta\,\ln{\frac{l}{\epsilon}} \label{eqn:main_E_tot_splay}
\end{align}
Similarly, the bend elastic energy is computed by substituting the bend director ($\hat{n}_{bend})$ from \eqref{eqn:ideal_n} in equation \eqref{eqn:frank_Oseen_main}. Following the same integration as described for the splay, we obtain the bend energy.
\begin{align}
E_{tot_{bend}}=(1-\frac{\pi}{\theta})^2\frac{\theta}{4}(k_1+k_3)  \ln{\frac{l}{\epsilon}}  \label{eqn:main_E_tot_bend}
\end{align}
\eqref{eqn:main_E_tot_bend} shows that in this case the energy depends on both bend and splay constants.

At equilibrium, the probabilities of the system being in either states can be calculated by taking the ratio of the Boltzmann weight ($P_i= e^{-E/k_BT}$). In Figure \ref{fig:Figure 2 Main}.E we can observe that the fraction of occurrence for the two deformations is roughly equal at $\theta=\frac{\pi}{2}$. Thus we take the formula for equilibrium and equate the the total energy for the two deformations given in \eqref{eqn:main_E_tot_splay} and \eqref{eqn:main_E_tot_bend}. This yields the following simple relation:
 \begin{gather}
      k_1 = k_3 \label{eqn:main_90D_constant}
 \end{gather}
This indicates that the assumption of the one-constant approximation commonly used for active nematics \cite{Perez-Gonzales2019,Doostmohammadi2018,Turiv2020,Saw2017,Comelles2021,Duclos2016,Wang2023,Hoffmann2020} holds also for 3T3 fibroblasts.

\begin{figure*}[ht!]
\centering
\includegraphics[width=\textwidth]{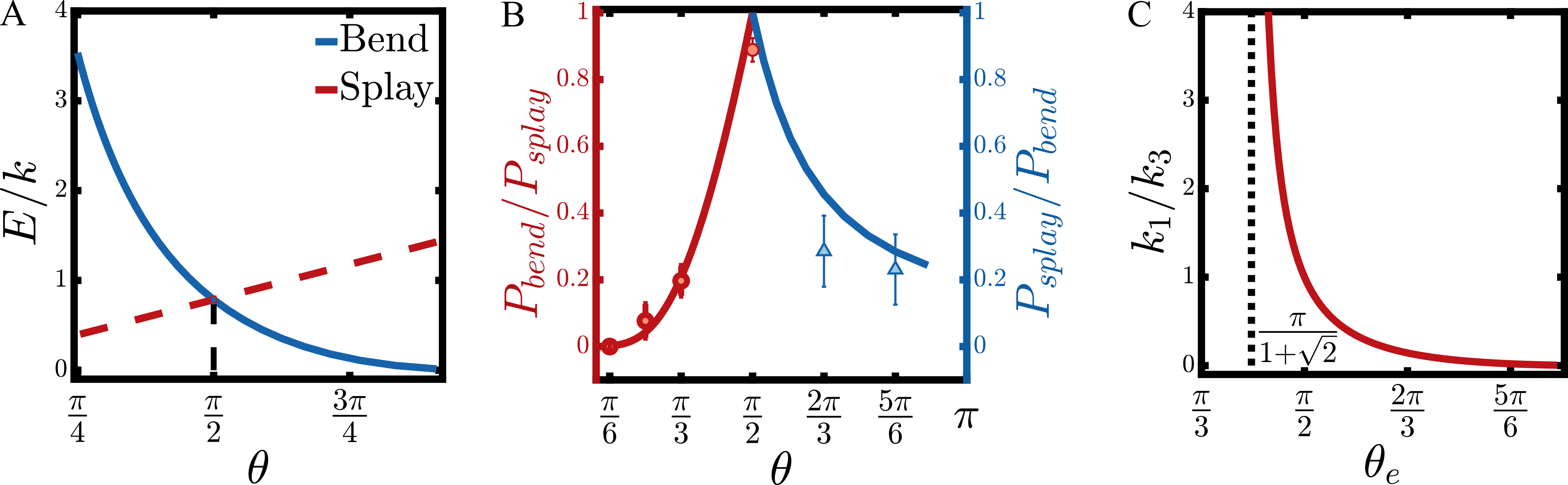}
\caption{(A) Total energy $(E)$ divided by the elastic constant as a function of $\theta$ under the assumption  $k_1\!=\!k_3$.  (B) The left axis (red) plots the ratio of bend over splay as a function of $\theta$ for theory (line) and for experiments (circles). The right axis (blue) of the graph plots the ratio of splay over bend as a function of $\theta$. (C) The ratio of splay and bend elastic constant as a function of $\theta_e$, i.e. the angle at which there is equal probability to splay or bend.} 
\label{fig:Figure 3 Main}
\end{figure*}

We can also consider the measurements at different angles to confirm our hypothesis. To explore this, we plot the energy divided by the elastic constant as a function of wedge angle ($\theta$), as shown in Fig.\ref{fig:Figure 3 Main}.A. The energy associated with splay (red line) grows linearly with $\theta$ \eqref{eqn:main_E_tot_splay}. In contrast, the energy for the bend case (blue line) exhibits a non-linear dependence, scaling as $\theta + \frac{c}{\theta}$ \eqref{eqn:main_E_tot_bend} where c is a constant, leading to a crossover point where the energies of splay and bend intersect. For angles less than $\theta = \frac{\pi}{2}$ (the transition point, black dashed line), splay represents the lower energy conformation compared to bend, with the roles reversing beyond this transition point. This theoretical result aligns well with our experimental observations, but we aim to take it a step further by directly calculating the probabilities from these energy values.

To calculate the relative probabilities, we use the Boltzmann weights ($e^{-E_{i}/k_BT}$) for splay and bend deformations.  When we take the ratio of the two probabilities, the ratio $e^{-E_{bend}/k_BT}/e^{-E_{splay}/k_BT}$ leaves a prefactor $\alpha=$ $\frac{\ln(l/\epsilon)k_1}{k_BT}$, which is unknown. For simplicity, we assume that this prefactor is equal 1. We should note that by fixing our measurement point at $l=325\mu$m, we can simplify our classification into only splay or bend and disregard the effect of splay to bend and bend to splay transition. With this hypothesis, the resulting probabilities are plotted in Fig.\ref{fig:Figure 3 Main}.B as solid lines (red and blue). The left axis shows the ratio $P_{bend}/P_{splay}$ (in red), with the corresponding theoretical probabilities plotted as a red line. For angles greater than $\theta = \frac{\pi}{2}$, where this ratio exceeds one, we instead plot $P_{splay}/P_{bend}$ (in blue) on the right axis to ensure the ratio lies between 0 and 1, with the associated line shown in blue. 

Our theoretical predictions are shown together with the experimental data, as the red circles in the figure indicate the ratio of the probability of bend over splay deformations from experiments for angles $\theta = \frac{\pi}{6}, \frac{\pi}{4}, \frac{\pi}{3},$ and $\frac{\pi}{2}$. Similarly, for angles greater than $\theta = \frac{\pi}{2}$, we plot the ratio of the probability of splay over bend deformations as blue triangles for angles $\theta = \frac{2\pi}{3}, \frac{5\pi}{6}$. The error bars are calculated by assuming the the distribution to be binomial with the number of trials specified in SI table S1. The qualitative good agreement between experimental and theoretical results supports the assumption that the splay and bend constants are equal as this hypothesis leads to a good data fit across all values of $\theta$.

It is possible that for other cell types or different substrates, the probabilities of the two deformations may not be equal at $\theta = \frac{\pi}{2}$, if $k_1 \neq k_3$. Even in such cases, we can calculate the elastic anisotropy ($k_1/k_3$) if we know the angle $\theta_e$ at which the probabilities of the two deformations are equal. By setting the probabilities of splay and bend equal for a specific value of $\theta_e$, we derive $k_1/k_3$ as a function of $\theta_e$:

\begin{equation}
 \frac{k_1}{k_3} =\frac{(1-\frac{\pi}{\theta_e})^2}{(1-\frac{\pi^2}{\theta_e^2}+2\frac{\pi}{\theta_e})} \label{eqn:main_elasticanisotropy}
\end{equation}

This function is plotted in Fig.\ref{fig:Figure 3 Main}.C. As $\theta_e$ decreases from $\pi$, the value of $k_1/k_3$ increases monotonically and diverges at $\theta_e = \frac{\pi}{1+\sqrt{2}}$. The equation also indicates that there exists a wide range of $\theta_e$ values, specifically $77^{\circ} < \theta_e < 126^{^o}$, where the two elastic constants are within an order of magnitude from each other, making the one constant approximation a good assumption. 

\section*{Discussion}
This work analyzes the alignment of fibroblast cells near a corner, showing their analogy to those observed in 2D nematic liquid crystals (Fig.\ref{fig:Figure 2 Main}.A). In analogy with liquid crystals, where alignment can be precisely controlled through patterning, we systematically study these deformations. Our findings indicate that cells undergo splay and bend-like deformations with a probability that depends on the wedge angle. The experimental data present an intuitive trend: as the wedge angle increases, cells transition from predominantly splay deformations near a corner to bend deformations. Notably, at $\theta = \pi/2$, the fractions of splay and bend deformations are nearly equal.

By calculating the Frank-Oseen elastic energies associated with these two deformations we determine that the two elastic constants must be equal to explain our experimental data. To further interpret the experimental results, we can calculate the relative probability of splay and bend configuration using the Boltzmann weights and compare it with the observations. 

This estimate however contains the prefactor $\alpha$ =$\frac{\ln(l/\epsilon)k_1}{k_BT}$ with unknown terms, namely the ratio between the observation length $l$  and the defect core size $\epsilon$ , and the ratio of the 2D elastic constant $k_1$ (due to the 2D nature of the problem, this constant does not have the usual units of force, but has the units of energy) and the thermal energy $k_B T$. It is reasonable to assume that the system can be equated to a thermal active bath with temperature $T$, and that the elastic constant will be equal to a few times $k_BT$. In nematic liquid crystals, the elastic constants are in the order of 10 pN, which gives an energy of $\approx4-5 k_BT$ if multiplied by the length of a typical mesogen. If we assume $k_1\approx4 k_BT$ , then $\alpha=1$ if  $l/\epsilon\approx$ 1.3, meaning the length at which we measure the configuration is between once and twice the defect core length. This makes physical sense. On the one hand, we want to test alignment over a lenght scale larger than the defect core size. On the other hand, measurements beyond the correlation length are meaningless because the alignment from the corner will be lost. So we are restricted on one side by the nematic coherence length, which we estimated from image analysis to be around 450$\mu$m. On the other hand we need to be far enough from the defect core, whose size we do not know exactly. Drawing again an analogy with molecular liquid crystal system, the defect core size should be the size of a few mesogens, so in this case a few cells. With a typical cell length around 50$\mu$m a defect core size around 150-200$\mu$m is a reasonable estimate. This estimate also finds confirmation in our previous work, where we estimated the length over which the cell density is altered due to the presence of a defect \cite{Endresen2021}.  Consistently with these boundaries, our measurements are taken at a length $l=315\mu$m from the corner.

This approach yields good agreement between experimental observations and theoretical predictions, even though the theory comes from equilibrium liquid crystal theory and equilibrium statistical mechanics. This crude approximation can still describe a non-equilibrium phenomenon with good qualitative matching.
In this way we can estimate the elastic anisotropy, the ratio of the elastic constants, as a function of the wedge angle $\theta_e$ in \eqref{eqn:main_elasticanisotropy}, which corresponds to the angle at which the probabilities of the two deformations are equal. This makes this method suitable for estimating the elastic anisotropy of a variety of cell systems, including those where the defect core is difficult to identify.

\section*{Experimental Methods}
\subsection*{Cell Culture}

The NIH-3T3 mouse fibroblast (from ATCC) are cultured in Thermofisher Scientific Nunclon$\mathrm{^{TM}}$ Delta Surface treated tissue culture dishes using 89 \% Dulbecco's Modified Eagle's Medium (DMEM) - high glucose [+] 4.5g/L glucose, L-glutamine, sodium pyruvate, and sodium bicarbonate (Sigma Aldrich), 10\% Fetal Bovine Serum (Sigma Aldrich), and 1\% Penicillin-Streptomycin. Cells used for experiments are between generations 11 and 16.

\subsection*{Substrate Manufacturing}
Templates for the master fabrication were made using SU-8 TF 6001 negative photoresist (Kayaku Advanced Materials) in a ISO 5 cleanroom facility. A 4´´ CZ-Si wafer was first cleaned by submerging it in piranha etch bath for 5 minutes, followed by de-ionized water rinse for 1 minute and then spin dried. SU-8 resist was then spin coated on the wafer, then the wafer was placed on a leveled hot plate at 110 °C for 2 minutes and 30 seconds. After cooling down the wafer was then exposed to the pattern using a MJB4 mask aligner (13 seconds at 8.3 mJ/cm$^2$). To complete the curing reaction a post exposure, bake step took place at 110$^\circ$C for 2 minutes and 30 seconds. The wafer then was immersed for 3 minutes in a petri dish filled with SU-8 developer (Kayaku Advanced Materials), agitation every 30 seconds allowed fresh developer to interact with the surface. Finally, the wafer was hard baked at 160$^\circ$C for 10 minutes. The structures were later inspected using a Nikon Eclipse LV100 microscope and a Veeco Dektak 150 stylus profilometer.

\subsection*{Substrate Preparation}
Polydimethylsiloxane (PDMS) from Sylgard 184 (Dow Corning) is mixed thoroughly with 10\% of curing agent. It is later desiccated at room temperature and then poured onto a negatively patterned SU-8 substrate. This is again desiccated to get rid of any leftover bubbles before being cured in a $60^o$ oven for 4 hr. The PDMS is peeled off the mold and the patterned PDMS is cut in slabs and used for making copy of substrate with UV-curable glue Norland optical adhesive 81 (NOA-81). The patterned substrate is flipped onto a flat-bottom glass petri dish with 2-3 drops of NOA-81  glue. The sample is degassed to make sure there are no bubbles in the glue below the patterned sample. The susbtrate is then radiated with UV at 302nm (8W, Ultra Violet Products-3UV) for 20 mins on each side. The petri dish is then heated at $60^{\circ} C$ for 30 mins after which the patterned substrate is removed and negative of the pattern is ready to be used as mold for making PDMS copies.

The patterned PDMS slabs are cleaned by rinsing with absolute ethanol and isopropyl alcohol and dried using compressed air. A thin layer (3g) of 1:10 ratio PDMS is poured on a petri dish and cured overnight in a $37^{\circ}$ oven. Both the petri dish and the patterned PDMS slabs are treated with oxygen plasma (Harrick Plasma Cleaner) with RF power 30W for 3 minutes with a pressure of 300mtorr. The PDMS slab is then attached to the coated petri dish with the substrate facing up and heat treated in a $60^{\circ}$ oven for 1 minute. The sample is plasma cleaned again with the same setting for 3 minutes and then heated for 1 minute at $60^{\circ}$.  

The sample is then sterilized with ethanol and it is prepped for substrate treatment with fibronectin. Fibronectin from bovine plasma (Sigma Aldrich) 25$\mu$g/ml solution is coated onto the substrate with minimal volume (200 $\mu$l for 3 cm$^2$ substrate area) for 45 mins at room temperature before being washed. The sample is then used for cell culture.

Cells are seeded onto the patterns, once the concentration of the suspension of cells is determined using Trypan Blue and counted using a hemacytometer (10 $\mu$l volume). The concentration of the cells used for plating in the petri dish is 500 cell/mm$^2$.

\subsection*{Staining}
To observe the nuclei the cells are stained using NucBlue$^\mathrm{TM}$ Live Cell ReadyProbes$^\mathrm{TM}$ (Hoescht 33342) by adding 2 drops/ml of the cell media, followed by a 30-minute incubation.

\subsection*{Microscopy}
Phase Contrast and fluorescent imaging in 2D is done using a Nikon Tl-Eclipse Widefield microscope using a Kinetix Scientific CMOS camera (Teledyne Photometrics). The patterned triangles are visualized by capturing large format images by translating a stage along a grid with 15\% overlap between the frames. The images are later stitched together using a Stitching (Grid/Collection stitching) plugin in ImageJ \cite{Preibisch2009}. 

\subsection*{Image Analysis}
Images obtained from the microscope are pre-processed by passing them through an ImageJ macro that combines the different channels and stitches them using the ImageJ Stitching Plugin \cite{Preibisch2009}. Once the images are stitched, the edges of the triangles are identified manually and then translated using a custom Matlab code. The code aligns all the triangles of the same type and makes one side of the triangle parallel to the x-axis (reference axis for calculating orientation). The triangle's vertices are then cropped over an area of $400 \mu m $ x $400 \mu m $. The cropped phase contrast image is used to get the orientation of cells using the OrientationJ plugin in ImageJ. OrientationJ window size of $ 13 \mu$m  is selected to get the average orientation of the cells over the $ 13 \mu$m x $ 13 \mu$m window. The orientation at each point is adjusted such that the reference is with respect to the x-axis. 

\subsection*{Splay-Bend Transition}
To characterize splay or bend at each corner, a circle is drawn centered at the vertex of the corner with a radius $r$ as shown in Fig.\ref{fig:Figure 2 Main}.B. The grid points (from OrientationJ calculation) closest to the circle are then selected and are dotted with the unit radial vector from the vertex to the grid point. The angle of each unit vector with respect to the x-axis of the image is taken to be $\Omega$. The dot product $\hat{r}\cdot\hat{n}$ is squared to limit the value between 0-1. In the case of splay deformation as the orientation at any given point is radially outwards from the point of the defect(corner), $ \{\hat{r}\cdot\hat{n}\}^2 $ will be 1 or close to 1, as seen in SI  Fig.3. As for bend, since the orientation is radial at 0 and $\frac{\pi}{2}$ and azimuthal for  $\Omega=\frac{\pi}{4}$, $\{\hat{r}\cdot\hat{n}\}^2$ will follow a characteristic curve going for 1 ($\Omega=0$) to 0 ($\Omega=\frac{\pi}{4}$) to 1 again at ($\Omega=\frac{\pi}{4}$) as can be seen in the Fig.\ref{fig:Figure 2 Main}.D.

\section*{Author Contributions}
AM conducted the experiments and in conjuction with FS developed the theoretical framework, AM and FS wrote the paper, KE and PA assisted AM in experiments, and AG was responsible for photolithography of the substrate.

\section*{Conflicts of interest}
There are no conflicts to declare.

\begin{acknowledgements}
We thank Junghoon Lee, and Konark Bisht for discussion. F.S. acknowledges funding from Novo Nordisk Foundation Recruit NNF21OC0065453.
\end{acknowledgements}


\bibliographystyle{apsrev4-1}

\clearpage

\onecolumngrid

\renewcommand*{\thefigure}{S\arabic{figure}}
\renewcommand*{\theequation}{S\arabic{equation}}

\makeatletter
\newcommand{\rmnum}[1]{\romannumeral #1}
\newcommand{\Rmnum}[1]{\expandafter\@slowromancap\romannumeral #1@}
\makeatother

\topmargin -.5in
\textheight 9in
\oddsidemargin -.25in
\evensidemargin -.25in
\textwidth 7in


\setcounter{equation}{0}
\setcounter{figure}{0}
\setcounter{table}{0}
\setcounter{section}{0}
\setcounter{page}{1}
\makeatletter
\renewcommand{\theequation}{S\arabic{equation}}
\renewcommand{\thefigure}{S\arabic{figure}}
\section*{Supplementary Information}

\section{Experiments}
\begin{figure}[h]
\centering
\includegraphics[width=\textwidth]{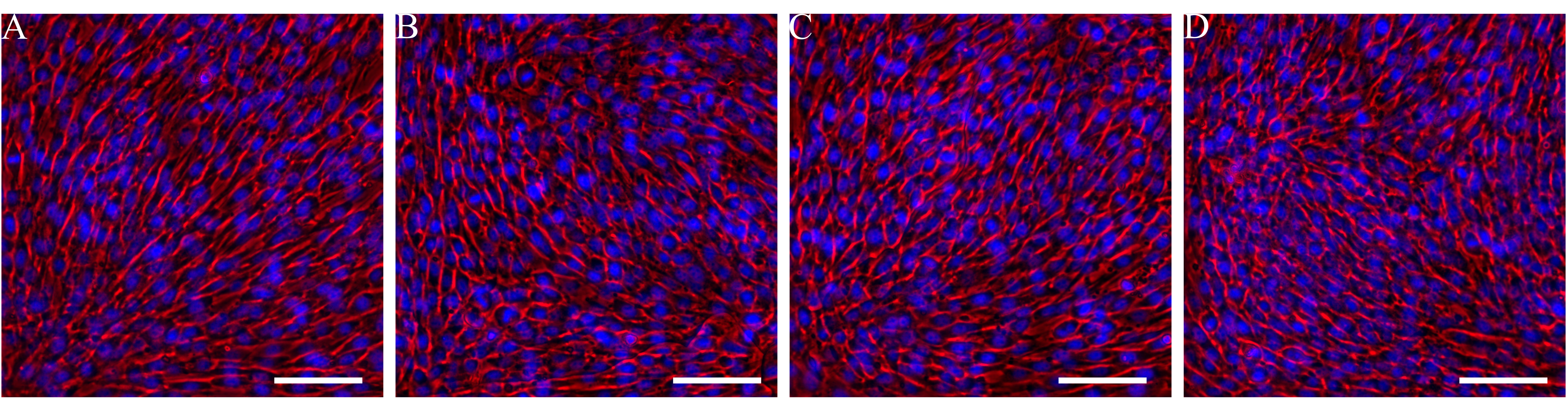}
\caption{Example of (A) splay (B) splay-bend (C) bend-splay (D) bend deformation that is observed in the experiments at $90^{\circ}$ vertex, positioned at the left bottom corner of each image. The red channel is the phase contrast imaging and the blue channel is the fluorescence imaging of cell nuclei stained with NucBlue LiveCell Stain for better visualization.(Scale Bar: 100 $\mu$m)}
\label{fig:SI_Figure_3}
\end{figure}

\begin{figure}[h]
\centering
\includegraphics[width=\textwidth]{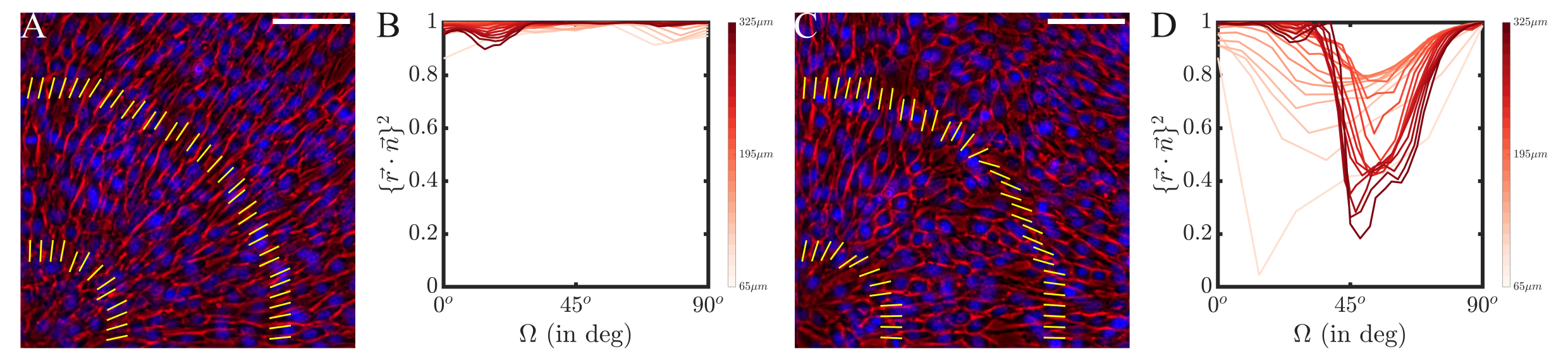}
\caption{(A) Representative example of a splay deformation, with two orientations shown at radial distances of 130 $\mu$m and 325 $\mu$m for deformations observed at a $90^{\circ}$ vertex, located at the lower left corner of each image. The red channel shows phase contrast imaging, while the blue channel displays fluorescence imaging of cell nuclei stained with NucBlue LiveCell Stain for enhanced visualization. (B) Plot of the function $\{\hat{r}\cdot\hat{n}\}^2$ vs. $\Omega$ for the splay deformation shown in A. (C) Example of an ``imperfect´´ splay deformation at radial distances of 130 $\mu$m and 325 $\mu$m, also observed at a $90^{\circ}$ vertex. (D) Plot of $\{\hat{r}\cdot\hat{n}\}^2$ vs. $\Omega$ for the deformation shown in C, which deviates from the expected splay trend but is still recognized as splay according to our categorization rule. (Scale Bar: 100 $\mu$m)}
\label{fig:SI_Figure_4}
\end{figure}

\begin{table}[ht]
\centering    
\begin{tabular}{lr}
Wedge angle $\theta$ & Number of data points \\
1. $\pi/6$ & 7 \\
2. $\pi/4$ & 28 \\
3. $\pi/3$ & 79 \\
4. $\pi/2$ & 21 \\
5. $2\pi/3$ & 18 \\
6. $5\pi/6$ & 16 \\
\end{tabular}
\caption{Statistics of different data points for various values of wedge angle ($\theta$)}
\label{Table:Number of Data for Each point.}
\end{table}

\section{Theory}
\subsection*{Splay and bend}
 In the Fig.2C in the main text the director field has strong planar anchoring with the edges of the triangle, and since the cells are arranged in a monolayer, the nematic director is confined to the $xy$ plane. We can thus describe the nematic director as $n_x = cos\,\phi$, $n_y = sin\,\phi$, and $n_z = 0$ from [46,47].

To describe the director orientation near the corners, we use the analytical form from [46]
\begin{equation}
 \phi=s\alpha + c \label{eqn:phi}
\end{equation}
where $\alpha= \tan^{-1}(y/x)$ and c is constant. The director field for the splay deformation, characterized by planar anchoring, can be conceptualized as a fraction of a +1 topological defect, with the angle going from $0$ to $\theta$ (wedge angle). Therefore $s=1$ and $c=0$ as per reference [46]. By inserting these values into \eqref{eqn:phi} and converting into cylindrical co-ordinates we get director field given by \eqref{eqn:SI_ideal_splay_director} for splay deformation near a corner.
For bend-like deformation, the liquid crystal molecules exhibit planar anchoring adjacent to the walls. The change occurs in between the walls, effectively shifting from radial alignment near the walls to azimuthal alignment in the middle. Therefore given the constraints we use a director orientation to be \eqref{eqn:ideal_bend_director} in cylindrical co-ordinates.
\begin{gather}
 \hat{n}_{splay}=\hat{r} + 0\hat{\phi}  \label{eqn:SI_ideal_splay_director}
 \\
 \hat{n}_{bend}=cos\, \left ( \frac{\pi}{\theta} \phi\right ) \,\hat{r} - sin\, \left(\frac{\pi}{\theta} \phi\right) \,\hat{\phi} 
 \label{eqn:ideal_bend_director}
\end{gather}

\begin{figure}[h]
\centering
\includegraphics[width=\textwidth]{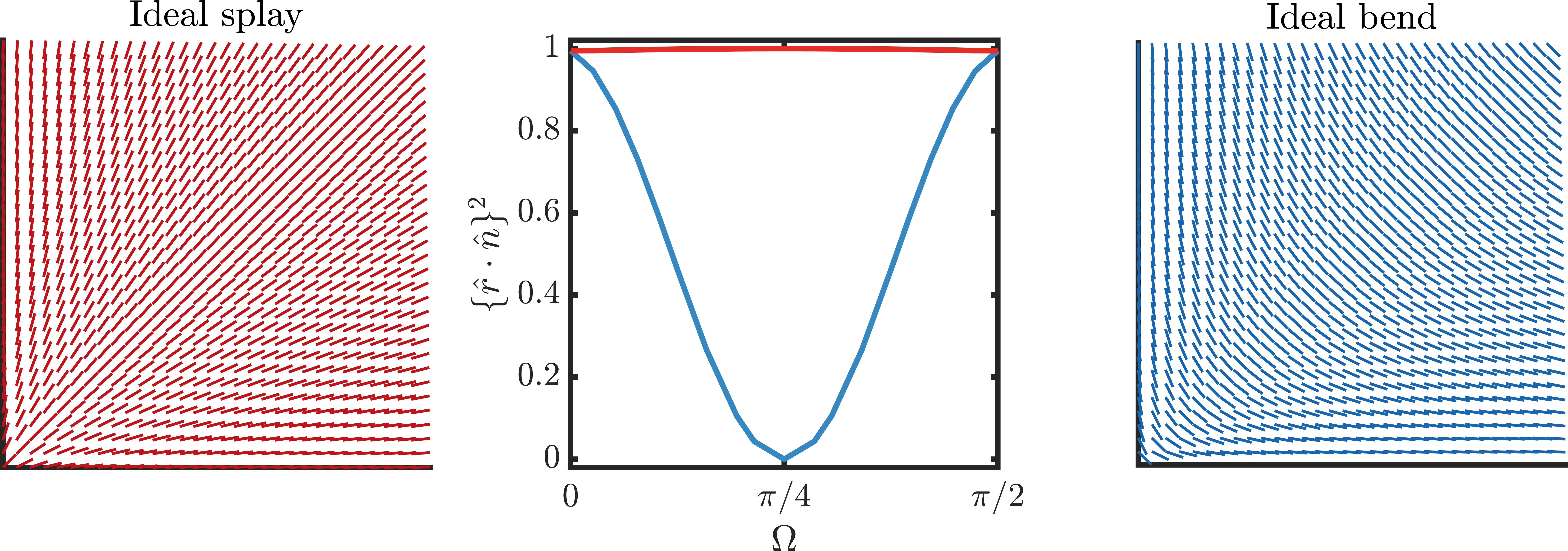}
\caption{Reconstructions of Splay (left, red) and Bend (right, blue) director field near a corner with amplitude $\pi/2$ and planar alignment on the edges. The central panel shows the values of $(\hat{r}.\hat{n})^2$ for the two calculated director fields.}
\label{fig:SI_Ideal_Condition}
\end{figure}

The plot of the value is given in Fig.\ref{fig:SI_Ideal_Condition} where we see the theoretical directors of the splay and bend plotted along with the analysis of how the $(\hat{r} \cdot \hat{n})^2$ varies as a function of $\Omega$

\begin{figure}[h]
\centering
\includegraphics[width=0.5\textwidth]{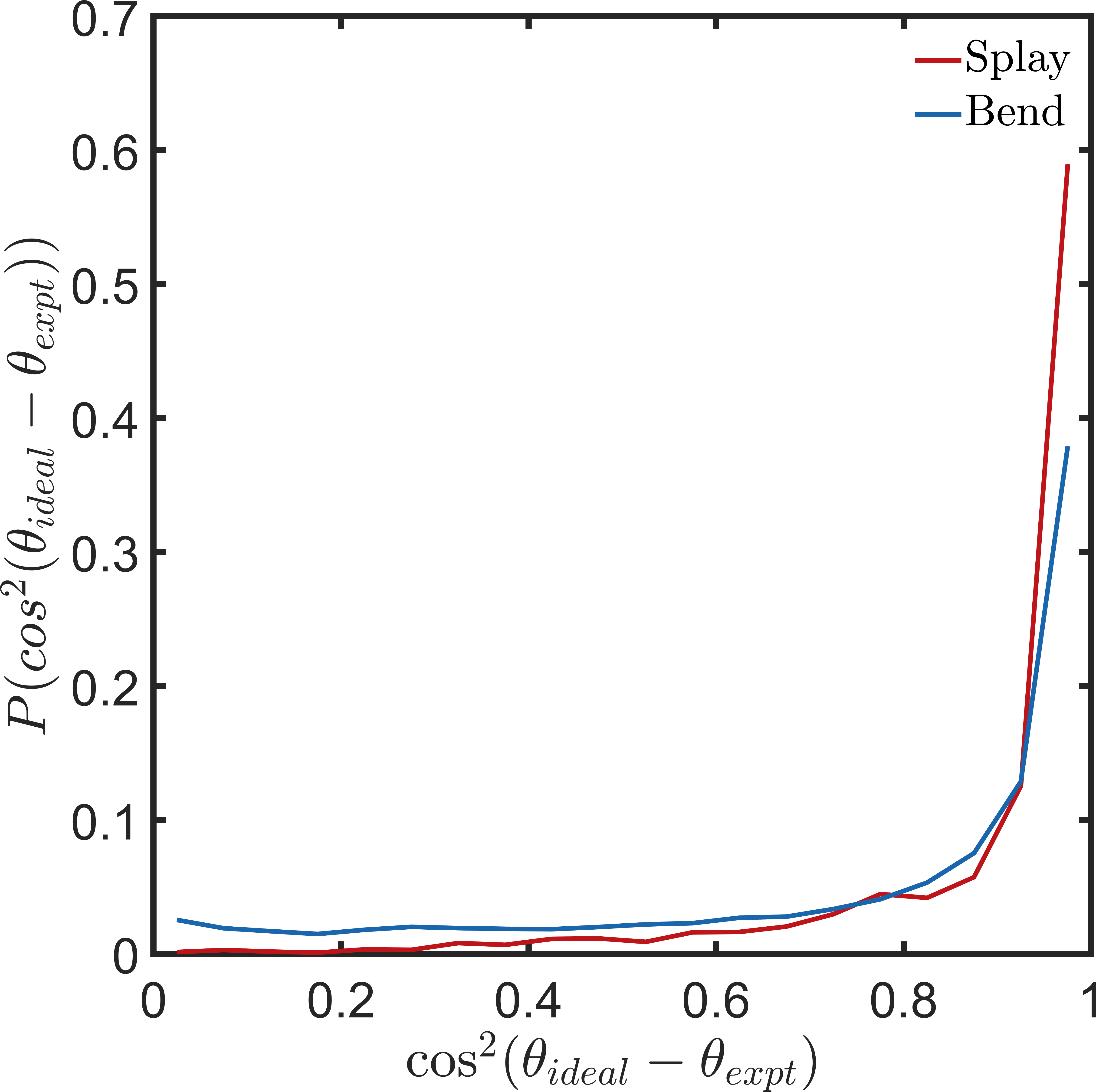}
\caption{Measured difference between the calculated director field and the experimental observations. $\theta_{expt}$ is the orientation of the director averaged over different experimental realizations for a $90^\circ$ wedge, for pure splay and pure bend case. $\theta_{ideal}$ is the theoretical director orientation for splay and bend deformation. In the figure the distribution of the difference between the experimental $\theta_{expt}$ and the theoretical $\theta_{ideal}$ for splay (red) and bend (blue) deformations is plotted as a function of the cosine square of $\theta_{ideal}-\theta_{expt}$}
\label{fig:SI_Figure_2}
\end{figure}

\subsection*{Energy Calculation}
The 3D Frank-Oseen energy is defined by \eqref{eqn:frank_onseen} [46,47] where $k_1$, $k_2$, and $k_3$ denote the elastic constants for splay, twist, and bend, respectively, and $\hat{n}$ represents the director field. As the cell monolayer is restricted to xy plane the twist energy term (characterized by $k_2$) from the \eqref{eqn:frank_onseen} can be discarded. Moreover the deformations can be considered only 2D, therefore both divergence and curl would be calculated only in 2D. This generates the \eqref{eqn:frank_onseen_2D} where the elastic constants are given by $k_1^*$ and $k_3^*$ which are the 2D splay and bend elastic constants. For simplicity, from now on we call them $k_1$ and $k_3$ even if they are defined in 2D. Integrating the Frank-Oseen energy over the wedge area and including the line anchoring term (2D analog to the surface anchoring energy) we get the total energy of the system \eqref{eqn:totalenergy}, where $W$ is anchoring energy per unit length, $l$ is the length of the wedge line, and $\hat{\tau}$ being the tangent vector to the edge. 

\begin{equation}
 f_{FO}=\frac{1}{2}k_1(\nabla\cdot{\hat{n}})^2+\frac{1}{2}k_2(\hat{n}\cdot\nabla\times{\hat{n}})^2+\frac{1}{2}k_3(\hat{n}\times(\nabla\times{\hat{n}}))^2  \label{eqn:frank_onseen} 
\end{equation}
\begin{equation}
 f_{FO_{2D}}=\frac{1}{2}k_1^*(\nabla\cdot{\hat{n}})^2+\frac{1}{2}k_3^*(\hat{n}\times(\nabla\times{\hat{n}}))^2  \label{eqn:frank_onseen_2D} 
 \end{equation}
\begin{equation}
 E_{tot}=\int_0^l \int_0^\theta f_{FO}\,r\,\partial \phi\,\partial r+2\int_0^l\,W(\hat{n}\times\hat{\tau})\,dl  \label{eqn:totalenergy} 
\end{equation}

As seen from Fig.2.C in the main text, the probability of cells aligning within 10 degrees of the wedge line is high for varying wedge angle. Thus we can imagine the line anchoring energy to be zero. We know from \eqref{eqn:SI_ideal_splay_director} the functional form of $\hat{n}$ for splay deformation. If we assume the wedge angle to be $\theta$ and substitute it in \eqref{eqn:frank_onseen} we get

\begin{gather} 
f_{FO_{splay}}=\frac{1}{2}k_1 \frac{1}{r^2} \label{eqn:SI_fo_splay} 
\end{gather}

because the bend term is zero. Substituting results of \eqref{eqn:SI_fo_splay} in  \eqref{eqn:totalenergy} we get the total energy for pure splay deformation as

\begin{align}   
E_{tot_{splay}} &=\int_\epsilon^l \int_0^\theta\frac{1}{2}k_1 \frac{1}{r^2}\,r\,d\phi\,dr \nonumber \\
&=\int_\epsilon^l \frac{1}{2}k_1 \frac{1}{r}\,\theta\,dr \nonumber \\
&=\frac{1}{2}k_1\theta\,\ln{\frac{l}{\epsilon}} \label{eqn:SI_E_tot_splay}
\end{align}

where $\epsilon$ is defect core, necessary to prevent the integral from diverging in zero. From \eqref{eqn:SI_E_tot_splay} you can see that the energy for splay is a linear function of $\theta$. One can do the same with the bend deformation case discussed before \eqref{eqn:ideal_bend_director} with $m\,=\frac{\pi}{\theta}$ and substitute in the \eqref{eqn:frank_onseen}

\begin{align}
f_{FO_{bend}}=\frac{1}{2}k_1 \frac{\cos^2{m\phi}}{r^2}(1-m)^2 +
\frac{1}{2}k_3 \frac{\sin^2{m\phi}}{r^2}(1-m)^2
\label{eqn:SI_fo_bend}
\end{align}

An important difference in this case as compared to the pure splay deformation is the presence of a component of energy coming from both splay and bend energy. Therefore the total energy from bend deformation is given by
\begin{align}
E_{tot_{bend}}&=\int_\epsilon^l \int_0^\theta \left[\frac{1}{2}k_1 \frac{\cos^2{m\phi}}{r^2}(1-m)^2+ \frac{1}{2}k_3 \frac{\sin^2{m\phi}}{r^2}(1-m)^2 \right] \,r\,d\phi\,dr \nonumber \\
&=\frac{1}{2}(1-m)^2\int_\epsilon^l\left[k_1 \frac{1}{r}\frac{\pi}{2m} + k_3 \frac{1}{r}\frac{\pi}{2m} \right]\,dr  \nonumber \\
&=\frac{(1-m)^2}{2} (k_1+k_3) \frac{\pi}{2m} \ln{\frac{l}{\epsilon}} \nonumber \\
&=(1-\frac{\pi}{\theta})^2\frac{\theta}{4}(k_1+k_3)  \ln{\frac{l}{\epsilon}}  \label{eqn_SI_E_tot_bend}
\end{align}
as $m=\pi/\theta$. It is of interest to note that the role of $\theta$ is not linear as seen with \eqref{eqn:SI_E_tot_splay}. We now have sufficient machinery to compare the two different deformations for the same angle.

\subsection*{Energy-Probability}
 The energy for the deformations are derived in \eqref{eqn:SI_E_tot_splay}, and \eqref{eqn_SI_E_tot_bend} for splay and bend respectively. We have seen from the experimental results, that for the case of fibronectin coated cell substrate shown in Fig.2E in the main paper, the probability to splay or bend is equal for the case when $\theta=\frac{\pi}{2}$. This suggests that the energy associated to splay and bend is equal, thus allowing us to equate \eqref{eqn:SI_E_tot_splay} and \eqref{eqn_SI_E_tot_bend} at $\theta=\frac{\pi}{2}$
 \begin{align}
 E_{splay} &=E_{bend} \quad \mathrm{when} \quad \theta=\frac{\pi}{2} \nonumber \\
 \frac{k_1}{2}\frac{\pi}{2}\,\ln{\frac{l}{\epsilon}} &=(1-\frac{\pi}{\frac{\pi}{2}})^2\frac{\pi}{2} \frac{(k_1+k_3)}{4}  \ln{\frac{l}{\epsilon}} \nonumber \\
 k_1 &= k_3 \label{eqn:SI_90D_constant}
\end{align}

\subsection*{Elastic Anisotropy}
The above condition from \eqref{eqn:SI_90D_constant} is valid for when the energy of splay and bend deformation are equal at $\theta=\frac{\pi}{2}$, what happens if that's not the case? We can still get the ratio between the elastic constants, using \eqref{eqn:SI_E_tot_splay}, and \eqref{eqn_SI_E_tot_bend}, and getting it as a function of $\theta_e$, i.e. the arbitrary angle at which the two energy are the same.

 \begin{align}
 E_{splay} (\theta_e) &=E_{bend}(\theta_e) \nonumber \\
 \frac{k_1}{2}\theta\,\ln{\frac{l}{\epsilon}} &=(1-\frac{\pi}{\theta})^2\theta \frac{(k_1+k_3)}{4}  \ln{\frac{l}{\epsilon}} \nonumber \\
 \frac{k_1}{k_3} &=\frac{(1-\frac{\pi}{\theta})^2}{(1-\frac{\pi^2}{\theta^2}+2\frac{\pi}{\theta})} \label{eqn:SI_elasticanisotropy}
\end{align}

the plot for this function vs $\theta_e$ is shown in Fig.3C of the main paper.

\end{document}